# Spin Transfer Torque Driven Coupled Oscillators for Self-Oscillating RF Mixers

Supriyo Maji
Department of Electrical & Computer Engineering, Purdue University,
West Lafayette, Indiana - 47907, USA

*Abstract*— Spin transfer torque oscillators (STOs) based on magnetic tunnel junction (MTJ) devices are emerging as a possible replacement for complementary metal-oxide semiconductors for radio-frequency (RF) signal generation. Advantages include low power consumption, small device area, and large frequency tunability. But such a single device cannot achieve the necessary noise performance for RF applications. It has been reported lately that a network of globally coupled STOs achieves significant improvement in phase noise. The study here is to propose use of such coupled STOs as self-oscillating RF mixers. Critical mixer performance parameters, including conversion gain, output power, and linearity, are discussed.

## I. Introduction

Spin transfer torque driven oscillation (STO) in magnetic tunnel junction device (MTJ) is emerging as a possible replacement of CMOS for RF signal generation [1], [2]. In addition to low power consumption and small footprint requirement, such oscillators can offer large frequency tunability over a wide range of input currents. In spite of several advantages, STO is rendered unacceptable for RF application due to its poor phase noise performance [3], [4]. In line with what was reported earlier in [5], that mutual injection of oscillators could produce better noise performance, [6] has proposed a globally connected network of spin torque oscillators to show substantial improvement in phase noise.

Mixer is an integral part of the up- and the down-conversion processes in RF transceiver. RF or baseband signal is mixed with the local oscillator signal for conversion into a suitable frequency range as part of the transmission or baseband processing. Some of the complexities of the conversion process can be eliminated by integrating oscillation and mixing operations [7]. Analytical solution has been proposed in [8] to characterize the physical phenomena of the frequency modulation or mixing observed before in STO devices [9]. Given inherent limitation of single STO device, it is proposed here that coupled oscillators as reported in [6], can be used for RF mixing to achieve better noise performance. Simulation study introduced here is based on the solution of Landau–Lifshitz–Gilbert –Slonczewski equation (LLGS, [10]) with macro-spin approximation.

The rest of the paper is organized as follows. Section II briefly discusses on the device properties and state of the art on phase noise performance of STO. Section III proposes RF mixing using coupled STOs and analyzes several critical performance parameters of the device. Finally, conclusions are drawn in section IV.

## II. Spin Transfer Torque Based Oscillation

Magnetization dynamics in MTJ device is described by LLGS equation [10], as shown below.

$$\frac{d\vec{m}}{dt} = -\gamma \left(\vec{m} \times \vec{H_{eff}}\right) - \alpha \left(\vec{m} \times \frac{d\vec{m}}{dt}\right) + \gamma \beta \varepsilon \left(\vec{m} \times \vec{m_p} \times \vec{m}\right) \quad \text{- (1)}$$

Here, $\vec{m}_x^2 + \vec{m}_y^2 + \vec{m}_z^2 = 1$ where $\vec{m}$ is the unit vector of magnetization, $\gamma$ is the gyromagnetic ratio, $\vec{H_{eff}}$ is the effective magnetic field, $\alpha$ is the damping parameter, $\varepsilon$ is the material dependent parameter, and $\beta$ is the function of DC bias current, volume ($V$) and saturation magnetization ($M_S$). Effective magnetic field constitutes of three main components: external ($\vec{H_{eff}}$), anisotropic ($\vec{H_{an}}$) and demagnetization ($\vec{H_{dem}}$).

$$\vec{H_{eff}} = \vec{H_{ext}} + \vec{H_{an}} + \vec{H_{dem}}, \quad \text{- (2)}$$
$$\vec{H_{an}} = K_1 \vec{m}, \quad \text{- (3)}$$
$$\vec{H_{dem}} = K_2 \vec{m} \quad \text{- (4)}$$

Where, $K_1$ and $K_2$ are respectively material dependent anisotropic and demagnetization constants. Since, operation of the device at no external field is desirable, study here is based on an in plane-perpendicular MTJ (Fig. 1(a)). Recently, it has been experimentally demonstrated that such devices can produce sustained oscillation without any external field [11]. A benchmarking with the experimental result has been reported in [6] and is used here for the simulation purpose.

### A. Phase noise

As with any oscillator, one of the most important figure of merit is phase noise. The first measurement in an STO showed a phase noise of -42dBc/Hz at 1 MHz offset which is too large for communication applications. Practical communication applications require a phase noise figure typically in the range of -100 dBc/Hz for 1 MHz offsets [3]. In MTJ, phase noise is caused by the device noise which leads to random deviation of the oscillator phase from that of an ideal reference oscillator. In addition to thermal noise in MTJ, there is low frequency flicker noise. Due to inherent mixing operation in a non-linear oscillator like STO, noise gets up-converted to the fundamental oscillation frequency. Flicker noise which has 1/f dependency appears as $1/f^3$ while thermal noise which has a constant power spectral density appears as $1/f^2$. Since, a transient representation of flicker noise is difficult to implement, only thermal noise has been considered here for the simulation.

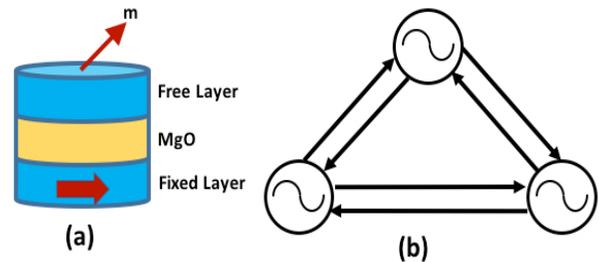

Figure 1. (a) Typical MTJ device (elliptical, in plane-perpendicular). (b) Global coupling of three oscillators.

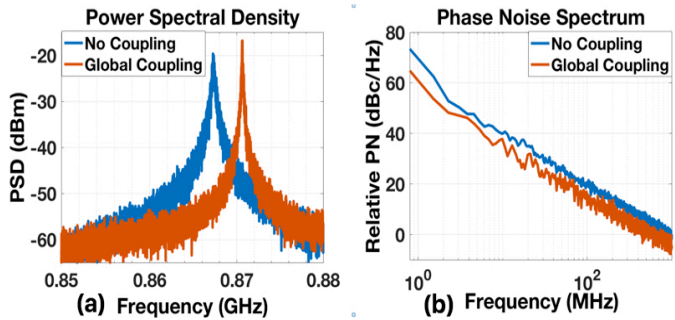

Figure 2. (a) Power spectral Density (PSD) plot for global and no coupling case. (b) Relative phase noise (PN) plot for global and no coupling involving three oscillators.

Previous studies have looked into material parameters and various device configurations which could improve phase noise performance of a single STO device [3], [4], but assessment of noise performance of a network of spin torque oscillators has been largely ignored. Similar to CMOS oscillators, STO is a non-linear device which conforms to general auto-oscillator theory [8]. Since, theoretically any non-linear oscillators can be coupled mutually or by injection locking for noise reduction [5], synchronization of STOs should also lead to improvement in phase noise performance. Recently, [6] proposed a globally connected network (Fig. 1(b)) of spin torque oscillators where significant improvement in phase noise over a single such devices has been observed through simulation. From the time-domain perspective, such synchronization or mutual coupling effect manifests itself as a correction of the oscillator zero crossings in every cycle which results in reduced accumulation of phase noise [12]. If more devices are coupled, noise performance should get better; though excessive amount of power and area requirements and the complexity in building CMOS interfacing circuits outweigh the benefit. A globally connected network of three oscillators yields optimal performance in terms of noise, power and area requirement [6]. Fig. 2 shows power spectral density and phase noise plots to compare coupling and no coupling behaviors. It is important to note here that the better phase noise figure can be attributed to the reduction in linewidth as well as the rise in the output power level.

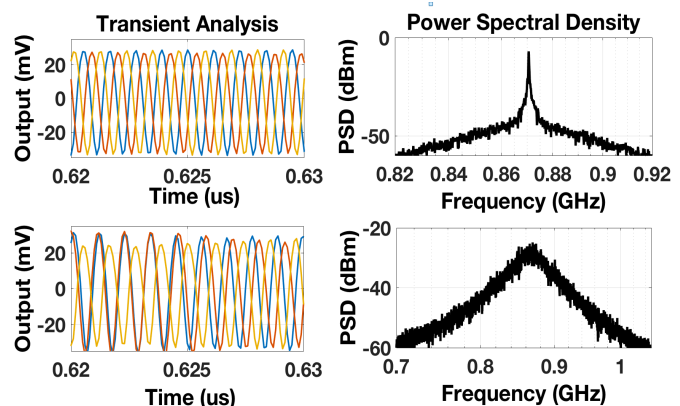

Figure 4. Timing simulation (left) and corresponding FFT (right) showing locking in phase and unlocking behavior of globally coupled three oscillators for V =1.6×150×70 nm$^3$(Top) & for V = 0.8×21×10 nm$^3$ (bottom).

### III. COUPLED STO BASED MIXER PERFORMANCE EVALUATION

A schematic diagram in Fig. 3 (a) shows how two oscillators are coupled and mixed with an RF input. A Verilog - A model to solve LLGS equation for spin torque oscillator has been developed for the purpose of calculating CMOS circuit requirements. Phase noise and power spectral density results have been obtained in Matlab. Since, STO is a current driven oscillator, a voltage to current conversion (V-I conv.) circuit is used to form the feedback loop. Control over the coupling strength happens through trans-conductance parameter of the PMOS transistor. Since DC current feedback would change the operating point or the frequency of operation, a capacitor is inserted in the feedback path to ensure that only AC current flows between the STOs.

Power spectral density of mixer output is plotted in Fig. 3(b) for both no coupling and global coupling. Since thermal noise is inversely related to the volume, effect of noise is more prominent in smaller devices [13]. This large thermal fluctuation can destabilize magnetization precession which in turn impedes synchronization effect between the oscillators. A timing simulation (Fig. 4 (left)) shows locking and unlocking behavior of the oscillators and the corresponding power spectral density is plotted in the same Fig. 4 (right).

#### A. Conversion gain, output power and linearity

Output power of STO depends on the DC bias current ($I_{dc}$), TMR and average resistance ($R_{av}$). With current ($I_{ac}$) as the input, RF input power ($P_{in}$) can be calculated by the expression, $P_{in} = (\frac{I_{ac}}{\sqrt{2}})^2 \times R_{av}$, Here, $R_{av} = (R_{AP} + R_P)/2$, $R_{AP}$ and $R_P$ are the MTJ anti-parallel and parallel resistances respectively. For output power calculation, FFT is performed on the output and side band power level is observed. Output power is estimated to be -39.83 dBm when sinusoidal input signal with a frequency of 300 MHz and a power of -11.1 dBm is injected for an oscillator frequency of 900 MHz. A conversion loss falls within 26.8 dB to 28.8 dB for this mixer for an input power ranging from -25.09 dBm to -11.11 dBm. Similar to the graphene based mixer where its conversion gain depends on the trans-conductance parameter of the transistor [14], the conversion gain and therefore the output power in STO can be improved with devices having larger TMR and average resistance.

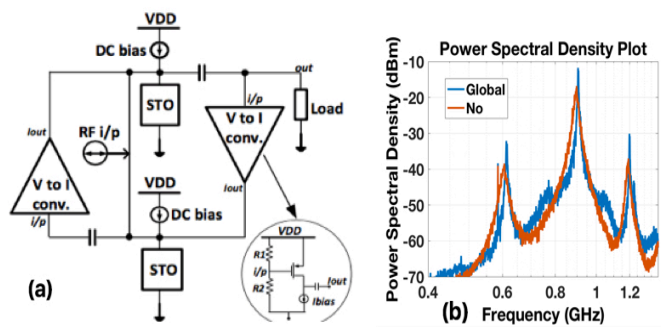

Figure 3. (a) Schematic showing coupling of two oscillators and RF current injection. (b) Mixer power spectral density plot with thermal noise for global coupling and no coupling.

TABLE - I
Performance Comparisons with Other Mixers

| Parameters | This work: STO mixer (*) | Graphene mixer [14] (**) | CMOS mixer [7] (**) |
|---|---|---|---|
| DC power consumption | STO: 0.6 mW CMOS: 1.2 mW at 1 V | -- | 5.4 mW |
| Frequency | 0.9 GHz | 10 MHz | 1.57 GHz |
| Volume or area | 50400 $nm^3$ | 1500 $\mu m^3$ | 1.5 $mm^2$ |
| Conversion gain | -28.72 dB | -30 dB | 36 dB |
| IIP3 | 8.9 dBm | 13.8 dBm | -19 dBm |
| 1dB compression point | -11.11 dBm | -- | -31 dBm |

*(*) Simulation results for three oscillators. (**) Fabrication results.*

Two important metrics determine the linearity performance of the non-linear device such as STO: 1 dB compression point and $3^{rd}$ order intercept point (IIP3). To measure the 1 dB compression point, input power is varied and power at the fundamental frequency component is observed. The 1 dB decrease can be specified at the input power level that produces it or the output power where 1 dB drop occurs. For the device under consideration, 1 dB compression occurs at -11.11 dBm of input power (Fig. 5(a)). For IIP3 measurement, higher order harmonics power is observed. For larger input power, the higher harmonics would start dominating and eventually equal the fundamental power (Fig. 5(b)). The point at which this occurs is called the IIP3 with respect to the input and OIP3 with respect to the output. IIP3 and OIP3 are measured to be 8.9 dBm and -21.6 dBm, respectively (Fig. 5(b)). A performance comparison with respect to the state of the art is presented in Table – I.

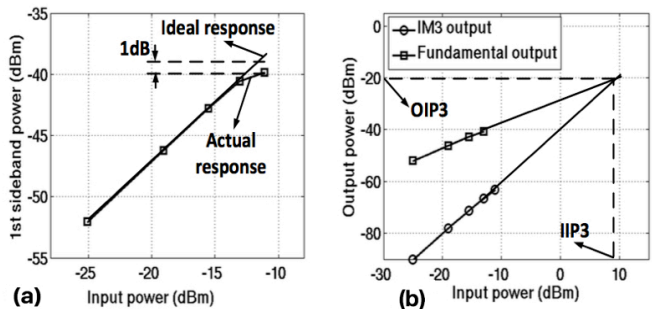

Figure 5. (a) Ideal and actual output power with increasing input power. (b) Fundamental and $3^{rd}$ order harmonic output power with increasing input power.

## IV. CONCLUSION

Spin transfer torque driven oscillation in nano-scale magnetic tunnel junction devices can be an attractive solution for RF signal generation with lower drive power, smaller footprint and larger tunability. It has been reported before that a globally connected network of spin torque oscillators can provide substantial improvement in phase noise. Here, in this paper, it is proposed that such coupled oscillator is useful for RF mixing. Though performance of the STO based self oscillating mixers appears to be better than other emerging devices, further improvement in output power and phase noise is necessary for demanding RF applications.